# AliEn – EDG Interoperability in ALICE

S. Bagnasco, P. Cerello
*I.N.F.N. Sezione di Torino, Torino, Italy*

R. Barbera
*Universita di Catania and I.N.F.N. Sezione di Catania, Catania, Italy*

P. Buncic, F. Carminati, P. Saiz
*CERN, Geneva, Switzerland*

AliEn (ALICE Environment) is a GRID-like system for large scale job submission and distributed data management developed and used in the context of ALICE, the CERN LHC heavy-ion experiment. With the aim of exploiting upcoming Grid resources to run AliEn-managed jobs and store the produced data, the problem of AliEn-EDG interoperability was addressed and an interface was designed. One or more EDG (European Data Grid) User Interface machines run the AliEn software suite (Cluster Monitor, Storage Element and Computing Element), and act as interface nodes between the systems. An EDG Resource Broker is seen by the AliEn server as a single Computing Element, while the EDG storage is seen by AliEn as a single, large Storage Element; files produced in EDG sites are registered in both the EDG Replica Catalogue and in the AliEn Data Catalogue, thus ensuring accessibility from both worlds. In fact, both registrations are required: the AliEn one is used for the data management, the EDG one to guarantee the integrity and access to EDG produced data. A prototype interface has been successfully deployed using the ALICE AliEn Server and the EDG and DataTAG Testbeds.

## 1. ALIEN AND THE EDG TESTBED

The ALICE collaboration has been using AliEn (ALIce ENvironment, a grid-like system for job submission and data management, described in detail elsewhere in this proceedings) as a production system for several Montecarlo simulation, reconstruction and analysis productions since the end of 2001 [1].

Based on existing, open-source modules, it is a simpler, lightweight alternative to full-blown Grid projects; its features include:

- Hierarchical, MySQL-based Data Catalogue with command-line and GUI interfaces.
- "Pull-model" for job submission, with built-in support for massive productions.
- LDAP-based remote configuration of sites.
- Support for metadata management.
- Web-based monitoring and management.
- Installation-on-demand of required software on remote sites.

In the near future, most of the available computing resource in the LHC community will be under the control of some flavour of Grid middleware (such as in the LCG Project); the aim of this work is to enable the ALICE collaboration to exploit seamlessly any computing resource (whether owned or shared, under the direct control of AliEn or seen through some Grid Resource Broker). This paper describes the interfaces of the AliEn system to the European DataGrid (EDG) and the European DataTAG projects testbeds [2,3].

Since full interoperability (i.e. the capability to exchange jobs between the systems in both directions) is not needed, an alternative *interfacing* approach has been devised, exploiting the modularity of the AliEn system: the whole Grid (namely, a Resource Broker) is seen by the server as a single AliEn Computing Element, and the whole Grid storage is seen as a single, large AliEn Storage Element.

In this architecture, the contact points between the two systems are the "Interface nodes": Grid User Interface machines that run also, as services, the AliEn client suite: Storage Element (SE), Computing Element (CE), Cluster Monitor and File Transfer Daemon. The interface CE gets jobs from the AliEn server and submits them to the EDG resource broker, taking also care to handle file registration and staging across the system boundary.

The advantages of such strategy are manifold:

- Ease of implementation: due to the object-oriented modular design of the AliEn code, the interface to the EDG Replica Catalogue is implemented as an instantiation of the "MSS" object (a Mass Storage System, like CASTOR or HPSS), and the interface to the Resource Broker as a "Local Queue" object (like PBS or Condor); the actual coding consisted in writing just two classes without any modification elsewhere in the system.
- For the same reason, it is possible to easily interface to different Grid middlewares; currently AliEn can submit jobs to "plain" EDG and to the DataTAG GLUE-compliant middleware.
- No need for services permanently running on remote sites; the interface nodes are the only ones running AliEn as a set of daemons, while on EDG Worker Nodes AliEn clients are started by the job itself.
- Ease of installation on EDG sites, since AliEn RPM packages are set up as any other application software, with no interference with the middleware.

The first such site has been installed in Torino, Italy, on a machine running EDG 1.4.3 and AliEn 1.29.9, submitting jobs to the EDG and DataTAG testbeds.

## 2. JOB SUBMISSION

As previously mentioned, the Job Submission interface is implemented as a "Local Queue"; a Grid Resource Broker is thus seen as a single queue.

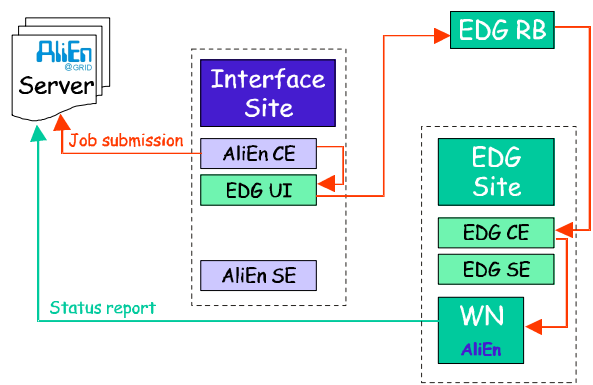

Figure 1: AliEn to EDG job submission.

When the AliEn Computing Element "pulls" a job from the server, a new JDL file is generated from the original AliEn JDL (which is not directly compatible with EDG JDL), translating the different syntaxes and adding:

- requirements for a suitable version of AliEn runtime environment;
- requirement for outgoing network connectivity from the Worker Node (needed by the AliEn code running on the WN to contact the server)
- environment variables containing the ProcID and the JobToken (used by the WN to authenticate itself with the server) and some configuration information needed to run AliEn on the WN (e.g. the address of the AliEn LDAP server)

In AliEn JDL files, a SpecialRequirements ClassAd attribute can be used to pass requirements directly to the Resource Broker that are not meaningful to AliEn.

At the same time, a job wrapper shell script is generated to set up some more environment (typically, configurations needed by the application software); the job is then submitted to the Resource Broker, passing the shell script via the InputSandbox.

The job wrapper script is needed since when submitting to EDG sites the AliEn Package Manager module (that takes care of setting up the environment for applications and of the installation-on-demand of needed software on remote sites) is disabled.

As soon as the job starts on the WN, AliEn code is run there to contact the server, piping the stdout and stderr streams to the server's logging system; it is then possible from the Web Portal to monitor the progress of the job or to kill it if needed. A barebone interface to the EDG Logging and Bookkeeping system is also provided to follow the job's progress through the various submission steps.

A database of EDG/AliEn JobID correspondences is kept by the interface CE to keep track of jobs that did not report yet to the server, to ease error handling and problem tracking.

## 3. FILE REGISTRATION

All files generated by a job running on an EDG site are doubly-registered in both (AliEn and EDG/Globus) Replica Catalogues, thus ensuring accessibility from both systems; again, the Storage Element/Replica Manager interface is implemented in AliEn as a "Mass Storage System" object (like CASTOR or HPSS).

Whenever an output file generated by an AliEn job running on some EDG site is to be stored and registered in the catalogue, the CloseSE (in EDG terms) is determined by AliEn code (running on the WN) from the .BrokerInfo file. The file is then saved in the selected SE and registered in EDG Replica Catalogue (uniqueness of the Logical File Name is enforced by AliEn).

An "intermediate" file name of the form EDG://<VO>/<LFN> is generated using the EDG Logical File Name and the Virtual Organization name to allow form multiple VO/RC. The file is finally registered in AliEn Data Catalogue using this "intermediate FN" as Physical File Name; the file is thus seen by AliEn as "stored in EDG", regardless of the physical location of the EDG Storage Element on which it resides.

It will be care of the Interface SE to query the EDG Replica Catalogue, whenever needed, to translate the AliEn LFN into a valid EDG Physical File Name.

## 4. DATA ACCESS

File accessibility is guaranteed in both directions, i.e. jobs running on either side of the interface can access data on either system. However, since the Resource Broker tries always to send the jobs where the data are, the basic case is also the most frequent.

### 4.1. The basic case

Whenever a job (in its JDL file) requires data stored in an EDG Storage Element, the AliEn job broker will at first try to assign it to the EDG interface Computing Element. The interface machine queries the AliEn Data Catalogue to translate the AliEn Logical File Name (given in AliEn JDL requirements) into the corresponding EDG Logical File Name and generates the appropriate InputData ClassAds attributes to be included in the generated JDL; that is then submitted to the Resource Broker. The RB will then take care to send the job wherever the needed data are available.

Once the job is started, AliEn code running on the Worker Node will query again the AliEn Data Catalogue, and the EDG one afterwards, to actually retrieve the data (see fig. 2). In this simple case, all file transfers are local to the site.

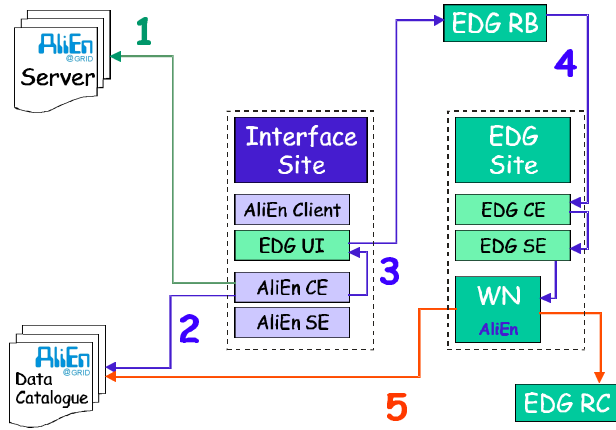

Figure 2: Data access – the basic case.

### 4.2. More complex occurrences

Should the need arise, it is possible for jobs running on either system to access data physically stored in the other; this would be the case, for example, for reconstruction jobs accessing large sets of data, partially residing in several different sites.

For example, access to data residing on EDG (but registered in the AliEn Data Catalogue) from AliEn sites, requires a two-step process: the job queries the AliEn Data Catalogue, and finds that the requested file is stored somewhere in EDG (that is all the information the AliEn Data Catalogue has).

It then asks the interface SE to retrieve the file. The interface asks the EDG Replica Manager to retrieve the file, and stages it on a temporary local area from which a File Transfer Daemon will move it to the requesting site.

The physical transfer of files is thus always performed by first staging them to the Interface machine, that again acts as a contact point between the two systems.

### 5.  OPEN ISSUES AND OUTLOOK

While the modular design of the AliEn system made the actual implementing of the interface code quite easy, a few problems were encountered.

The main one is that, unlike simple batch queue or MSS management software, the Grid is a complex system in which many different things can possibly go wrong. It is then one of the major tasks for the interface to gracefully recover from errors returned by some EDG call. Since the implementation was done using the Command Line Interface to EDG (AliEn is a set of Perl modules, and there are currently no Perl API's for EDG), the program has often to parse many lines of output from the commands to find out what the problem is; and that is often awkward.

For example, since many EDG commands are simply wrappers around lower-level software (e.g. the Globus toolkit), the format of the error message may be different if the error is generated by the wrapper script or just propagated from the underlying Globus command.

One of the main limitations of this approach is that, since all queries and data transfer must go through the interface site, it may become a bottleneck. A possible solution is to deploy more than one such nodes, each acting as a separate Computing Element requesting jobs from the server and submitting them to one or more Resource Brokers.

Another possible issue is that outbound connectivity is required from the Worker Node, in order to enable the job to report to the logging server. While this is not an issue in the current configuration of the testbed, since all nodes have public IP addresses; it may become a problem if in future Grid deployments the Worker Nodes will be hidden behind the Computing Element in closed clusters.

The interface site in Torino is currently being tested; while the code was not yet stable enough to be used in the Montecarlo production performed on the EDG testbed, the files produced have been subsequently registered in the AliEn Data Catalogue and are therefore available through AliEn.

It should be relatively easy to follow the future developments of the middleware; the migration from "plain" EDG JDL to the GLUE-compliant one required the modification of just few lines of code, and the release of EDG 2.0 should make things easier.

The plan for ALICE is to try to keep AliEn as the front-end for productions, sharing transparently the load between resources under the direct control of AliEn and, through the interface system, resources from the upcoming LCG testbed.

### Acknowledgements

This work was carried out as one of the tasks of Work Package 8 of the European DataTAG project.